\documentclass[10pt,aps,prb,onecolumn,nofootinbib,notitlepage,showkeys]{revtex4-2}

\usepackage{amssymb,amsmath}
\usepackage{subfig}
\usepackage{graphicx}
\usepackage{lipsum}
\usepackage{mathtools}
\usepackage{color}
\usepackage{hyperref}
\usepackage{bbm}
\usepackage{listings}
\usepackage{bm}
\usepackage{tikz}
\usetikzlibrary{arrows.meta}
\hypersetup{
    unicode=false,          % non-Latin characters in Acrobat’s bookmarks
    pdftoolbar=true,        % show Acrobat’s toolbar?
    pdfmenubar=true,        % show Acrobat’s menu?
    pdffitwindow=false,     % window fit to page when opened
    pdfstartview={FitH},    % fits the width of the page to the window
    pdfauthor={Kevin Multani},      % author
    colorlinks=true,        % false: boxed links; true: colored links
    linkcolor=blue,         % color of internal links
    citecolor=red,          % color of links to bibliography
}

\begin{document}

\title{Hypercomplex Yang-Mills Theory as a
Bipartite Gauge Field  Model}

\author{C. M. López Arellano}
\email{cmlopez@ifuap.buap.mx}

\author{R. Cartas-Fuentevilla}
%\email{rcartas@ifuap.buap.mx} % Opcional

\affiliation{Instituto de Física, Benemérita Universidad Autónoma de Puebla, Apartado Postal J-48 72570, Puebla Pue., Mexico}

\date{\today}

\begin{abstract}
A non-Abelian gauge field framework is proposed using the hypercomplex ring formalism. This extension generates non-compact hyperbolic symmetries, which, alongside the compact gauge symmetries, double the internal degrees of freedom. This will enable the description of bipartite gauge systems and demonstrate how field dissipation operates at the dynamical level. Working within a commutative ring allows for the decoupling of the algebraic structures and facilitates the construction of solutions to the equations of motion.
\end{abstract}


\maketitle
\section{Introduction}
\label{introduction}

The Yang-Mills theory developed in the 1950s \cite{Yang} has allowed us to study the dynamics at nuclear levels, revealing new interactions between elementary particles, such as the weak and strong forces, which along with electromagnetism allow classifying and describing much of observable matter. Starting from local gauge invariance principles, the Yang-Mills description generates fields which are associated with mediating particles of the forces \cite{Peskin}; it is here that the idea of bosons as mediators of fundamental forces emerges. This has allowed the construction of the standard model of particle physics, which is currently the most experimentally successful theory. 

Although the theory has agreed with experimentation, the standard model lacks the ability to explain thermodynamic phenomena such as dissipation; it is in this context that Thermo Field Dynamics (TFD) arises, seeking to relate zero-temperature quantum field theory with statistical mechanics, thus obtaining a framework for interactions at finite temperature \cite{Khan}. A characteristic of this formalism is the doubling of the degrees of freedom, which implies a doubling in the algebraic structure of the theory \cite{Santa}. The notion of doubled fields is not exclusive to thermal field scheme; theories such as twin Higgs models \cite{Chac} or extensions of the minimal standard model (MSM) \cite{Foot} start from this idea; this fact implies that models fulfill more general symmetries or that gauge fields can propagate over extra dimensions \cite{Appe}. 

In order to generalize the Yang-Mills formulation, the proposal in this letter is to extend our algebraic field over which the fields are defined. Here the idea of split-complex structures arises, which is an extension of real numbers, where a new complex unit $j$ appears, with the properties $j^2=1$ and $\bar{j}=-j$, \cite{Sob,Luna}. This new set $\mathbb{H} \equiv \mathbb{R}[j]$, which is a commutative ring, has been studied from a physical perspective, since Minkowski geometry can be modeled with this structure \cite{Naber}; these hyperbolic symmetries also been explored in gravitational models \cite{Zhong01,Zhong02}. By further enriching this set and incorporating the usual complex numbers $\mathbb{C} \equiv \mathbb{R}[i]$, a structure called hypercomplex numbers is obtained. Works using this approach have been developed; specifically, S. Ulrych has studied the hermicity of the Poincaré mass operator \cite{Url02} and representations of the Maxwell and Dirac equations, within a sixteen-dimensional structure containing internal and Lorentz space-time symmetries \cite{Url03}. Similarly, gauge theories for the Abelian case have been studied \cite{Car06}, where new properties have been found in scenarios of spontaneous symmetry breaking and specifically string-like anomalies have been constructed as cosmic topological defects in the presence of Aharonov-Bohm effects. 

The first indications for a hypercomplex formulation of non-Abelian gauge theories were explored also by S. Ulrych \cite{Url05}, where he outlines the scheme of Lie groups over hypercomplex numbers. Although he lays the foundations, he does not carry out a deep exploration, and his conclusions suggest that this formulation can give rise to new interactions. Motivated by these results, recent researches have found a physical interpretation from constructions of hypercomplex field theories for the description of dissipative systems \cite{Car03,Car01,Car02}, where it is not necessary to double the degrees of freedom due to some imposition; the formalism itself allows interpreting two complex fields $(\psi,\phi)$ as the dynamical variables of the theory, where we associate $\psi$ as the (sub-)system of interest and $\phi$ as the environment (thermal bath) acting as a mirror image evolving in a reverse time direction \cite{Car03}. This has led to quantization processes of dissipative systems, showing the existence of entangled state between the subsystem and environment. Basic aspects of ergodicity have been also explaned \cite{Car02}. A direct comparison with standard dissipation quantization has shown important differences \cite{Cele,Tra}. 

Following these line of thought, previous works have been built on scalar fields with global symmetry $\text{U}(1) \times \text{SO}(1,1)$; then our objective is to perform the generalization for gauge field theories with non-Abelian symmetry, that is, a hypercomplex dissipative Yang-Mills model. Recalling that the standard approach of this theory is formulated on fields and transformation groups over the field of complex numbers, thus describing compact symmetries \cite{Peskin,Tong}, which are represented by the special unitary groups SU$(n)$. We will show here that the presence of the hyperbolic contribution generates a new sector of non-compact symmetries, thus obtaining a scheme that unifies both symmetries.

We will work with an extended pure Yang-Mills theory, thus discarding coupling with matter, which is planned for a future work. This would allow us to explore the gluon-gluon interaction at a dissipative level, and would be the starting point to investigate quark-gluon plasma dynamics and transport phenomena at high energies \cite{Car03}, which has been studied through the AdS/CFT correspondence \cite{Poli}.  

In the next section, the scheme of hypercomplex numbers will be briefly developed; only the properties to be used in this approach will be shown. We will discuss how hyperbolic, and standard complex numbers are contained, revealing the underlying symmetries, and how the associated symmetry groups act on the subset of Hermitian numbers, which correspond to the generalization of real numbers; these objects will be fundamental since they will allow us to construct invariant quantities. In section \ref{sec:3} we will develop the theory of Lie groups over rings, where the group elements will be matrices with hypercomplex entries; this will lead to a division of the generators into two sectors, one compact and other non-compact; additionally it will be exemplified for internal dimensions $n=2,3$, for showing explicit the extensions for SU$(2)$ and  SU$(3)$ cases. For the extended algebra, in section \ref{sec:4} the structure of the covariant derivative and gauge fields are introduced, thus constructing the curvature tensor of the hypercomplex Yang-Mills; this framework can be associated with a subsystem/environment full system. Concluding with the derivation of the equations of motion of the theory which will show the interaction and dynamical mixing, namely, subsystem-reservoir gauge fields.  Remarks with future explorations are addressed. 

\section{The hypercomplex formalism}
\label{sec:2}

We define the commutative ring of hypercomplex numbers as follows \cite{Url03,Car05};
\begin{equation}
    z=x+iy+jv+ijw, \quad \bar{z}=x-iy-jv+ijw ; \quad x,y,v,w \in \mathbb{R}, 
    \label{01}
\end{equation}
where $i$ is the standard imaginary unit. Similarly, we introduce a new hyperbolic unit $j$ with the properties $j^2=1$, and $\bar{j}=-j$, along with a hybrid unit $ij=ji$ with the properties $\overline{(ij)}=ij$ and $(ij)^{2}=-1$. Note that the new conjugation on $z$ acts on both, the $i$ and $j$ units \cite{Url03}. If the entries variables $v=w=0$ are fixed in expression (\ref{01}), then $z\in \mathbb{C}$. Given the choice $y=w=0$, we obtain numbers of the form $z=x+jv,$ known as pure hyperbolic numbers $\mathbb{H}$ \cite{Sob}. This new algebraic structure $\mathbb{H}$ forms a commutative ring with a algebra similar to the usual complex numbers; however, the norm is not positive-definite, $z\bar{z}=x^2-v^2$. Observe that this quantity is invariant under continuous hyperbolic rotations, which can be represented by the connected component of the Lie group SO$(1,1)$ through the element $e^{j\chi}=\cosh{\chi}+j \sinh{\chi}$, where $\chi$ is a real non-compact parameter, and includes the identity element $I$ with $\chi=0$ \cite{Car06}, where the transformation follows the rule $z\rightarrow e^{j\chi}z$. The inverse of a hyperbolic number is given by $z^{-1}=\frac{x-jv}{x^2-v^2}$, with $x \neq \pm v$. Unlike complex numbers, the ring $\mathbb{H}$ contains two non-trivial idempotent forms, 
\begin{equation}
\begin{split}
J^{+}& \equiv \dfrac{1}{2}(1+j), \quad (J^{+})^{n}=J^{+}, \\
J^{-}& \equiv \dfrac{1}{2}(1-j), \quad (J^{-})^{n}=J^{-}, \quad n=1,2,3,\ldots;
\end{split}
\label{02}
\end{equation}
with the properties
\begin{equation}
    J^{+}J^{-}=0, \quad J^{+}+J^{-}=1;
    \label{proJ}
\end{equation}
  therefore $\mathbb{H}$ is not an integral domain \cite{Car05}. Other interesting properties of idempotent numbers include their action on the unit $j$, as shown in 
\begin{equation}
    J^{+}j=J^{+}, \quad J^{-}j=-J^{-}, \quad \overline{J^{+}}=J^{-}.
    \label{03}
\end{equation}
Note that $J^{\pm}$ acts on $j$ in such a way that $j$ is absorbed; this properties will be of great utility in the formalism developed here. 

From the above, it follows that the complex numbers $\mathbb{C}$ and the hyperbolic numbers $\mathbb{H}$ are contained in form (\ref{01}), where we denote the set of hyper-complex numbers as $\mathbb{H_C}$. Therefore, the norm  takes the form:
\begin{equation}
    z \bar{z}=x^2+y^2-v^2-w^2+2ij(xw-yv),
    \label{04}
\end{equation}
where it is observed that it maintains the structure of the norm of both a complex and a hyperbolic number, with the addition of a hybrid term $ij$. Thus, the norm of a hyper-complex number is no longer a real quantity; generally, we will refer to it as a Hermitian number, which is invariant under the action of circular rotations $e^{i\theta}$ represented by the Lie group $\text{U}(1)$, and the aforementioned hyperbolic rotations. Consequently, the quadratic form (\ref{04}) is invariant under the total rotation $z \rightarrow e^{i \theta}e^{j \chi}z$, which corresponds to $\text{U}\times \text{SO}(1,1)$ \cite{Car05}; this is relevant because, by constructing physical theories on Hermitian quantities, the non-compact symmetry $\text{SO}(1,1)$ will be incorporated into models together with standar compact internal symmetries \cite{Car06, Url05}.

The objects (\ref{02}) and (\ref{03}) defined in $\mathbb{H}$ are contained in $\mathbb{H_C}$. Noting that the product of the idempotents (\ref{02}) is null, it can be visualized as an orthogonality relation, allowing them to be used as projectors over hyper-complex numbers:
\begin{equation}
    J^{+}z=J^{+}z_{+}, \quad J^{-}z=J^{-}z_{-}, \quad z=J^{+}z_{+}+J^{-}z_{-};
    \label{05}
\end{equation}
where $z_{\pm}=x \pm y +i(v \pm w)\in \mathbb{C}$; thus we identify $(J^{+},J^{-})$ as a basis for $\mathbb{H_C}$, which allows us to rewrite hyperbolic rotations and exponential forms as follows:
\begin{equation}
    e^{j \chi}=J^{+}e^{\chi}+J^{-}e^{-\chi}, \quad e^{z}=J^{+}e^{z_{+}}+J^{-}e^{z_{-}}.
    \label{06}
\end{equation}
A Hermitian number can be written as $z_{0}=u_{0}+k w_{0}$, where $u_{0},w_{0}\in \mathbb{R}$ such that we have defined $ij=k$, with the properties $\bar{k}=k$ and $k^2=-1$. This form constitutes a closed subset denoted by $Z_{0}\subset \mathbb{H_C}$ with respect to addition and multiplication \cite{Car01}. Its inverse can be defined as:
\begin{equation}
     z_{0}=u_{0}+kw_{0}, \quad z_{0}z_{0}^{-1}=z_{0}^{-1}z_{0}=1, \quad z_{0}^{-1}=\dfrac{u_{0}-ij w_{0}}{u_{0}^2+w_{0}^2};
    \label{07}
\end{equation}
other properties and a polar representation of Hermitian forms can be found in \cite{Car05}.

\section{Hypercomplex formulation for unitary groups}
\label{sec:3}

Recalling that the Yang-Mills SU(2) formulation, through the idea of local non-abelian continuous transformations, attempted to describe the isotopic spin for the nucleon \cite{Yang}, which in the first years after its publication was not taken into account. It was R. Utiyama who axiomatized this model for any set of transformations dependent on $n$ parameters \cite{Uti}. The SU$(3)$ quark model proposed by Gell-Mann and Zweig provided a systematic framework for hadron classification \cite{Grei}, thereby establishing the theoretical underpinning of the Standard Model, which is based on the $\text{SU}(3) \times \text{SU}(2) \times \text{U}(1)$ gauge group and describes all known fundamental particles. 

As mentioned in the introduction, S. Ulrych in \cite{Url05} provides a brief introduction to the scheme of unitary groups over the hypercomplex ring; therefore, in this section we will develop these ideas in detail. It should be noted that works on unitary groups over the set of hyperbolic numbers $\mathbb{H}$ have been reported, where irreducible representations in this scheme are explored \cite{Popov01,Popov02}. 

\subsection{Lie Groups over commutative rings}
\label{sec:3.1}
Representations for SU$(n)$ groups are given by $n \times n$ matrices with complex entries; the objective of this work is to generalize these entries using hyper-complex numbers of the form (\ref{01}). We begin with the definition of the unitary group on $\mathbb{H_C}$,
\begin{equation}
    \text{U}(n,\mathbb{H_C})=\{ U \in \mathbb{H_C}(n) \mid \hspace{0.1cm} U \cdot {U^{\ddagger}}=\mathbbm{1} \},
    \label{08}
\end{equation}
where $\mathbb{H_C}(n)$ denotes the set of $n \times n$ matrices with hyper-complex entries. Note that we have used the involution convention\footnote{Let $A_{ij}\in \mathbb{H_C}$ be elements of a matrix $A$, the matrix involution is defined as $(A^{\ddagger})_{ij}=\overline{A_{ji}}$.} where $i$ and $j$ units change of sign \cite{Url05}. An exponential representation of (\ref{08}) can be given as $U=e^{iH}\in \mathbb{H_C}$; for the unitarity condition $ U \cdot {U^{\ddagger}}=\mathbbm{1}$ to be satisfied, the matrix $H$ must be Hermitian under the conditions \cite{Grei},
\begin{equation}
    H_{ii}=\overline{H_{ii}}, \quad H_{ij}=\overline{H_{ji}}, \quad i,j=1,2,\ldots,n,
    \label{09}
\end{equation}
where the diagonal elements can only be Hermitian numbers of the form (\ref{07}). By inspection we realize that both $H$ and $U$ have $2n^2$ independent parameters. Again, the unitarity condition provides a property for the determinant:
\begin{equation}
   \lvert \det U \rvert^2=1, \: \det U=\det e^{iH}= e^{i \text{Tr}H}=e^{i(\alpha+ij\beta)}; \: \alpha,\beta \in \mathbb{R}
    \label{10}
\end{equation}
this is because the trace of $H$ is a Hermitian number of the form (\ref{07}), unlike the case of SU$(n)$, where the trace is a real quantity \cite{Grei}. 
This restriction on the determinant leads to define a new group, with elements having unit determinant:
\begin{equation}
    \text{SU}(n,\mathbb{H_C})=\{ U \in \text{U}(n,\mathbb{H_C}) \mid \hspace{0.1cm} \det U={1} \},
    \label{11}
\end{equation}
where this new condition on the group element induces the traceless restriction $\text{Tr}\hspace{0.1cm}{H}=0$ such that $\alpha=\beta=0$. These elements will be described by $2(n^2-1)$ real parameters. This agrees with the field doubling idea, where there will be a sector associated with the standard SU$(n)$ with its respective $n^2-1$ compact generators and parameters; on the other hand, the same number of non-compact generators will be generated due to the hyper-complex extension \cite{Url05}. Then, the Hermitian matrix $H$ is broken down into two sectors as follows 
\begin{equation}
    H=\sum_{a=1}^{n^2-1}\theta^{a}t_{a}+\sum_{a=1}^{n^2-1}\chi^{a}\pi_{a}=\sum_{a=1}^{n^2-1}\bm({\alpha}^{a})^{T}\bm{X}_{a}, %\quad \bm{\alpha}^{T}_{a}=( \theta_{a} \hspace{0.2cm} \chi_{a}), \quad \text{and} \quad  \bm{X}_{a}=\begin{pmatrix} t_a \\ \pi_a \end{pmatrix}. 
    \label{20}
\end{equation}
where $\bm({\alpha}^{a})^{T}=( \theta^{a} \hspace{0.2cm} \chi^{a}),$ and $\bm{X}_{a}=\begin{pmatrix} t_a \\ \pi_a \end{pmatrix},$ such that the real parameters $\theta_a$ correspond to circular rotations and $\chi_a$ to hyperbolic rotations. Therefore, the unitarity restriction and unit determinant (\ref{11}) characterizes the set of generators, which allows us to define the following algebra,
\begin{equation}
    \mathfrak{su}(n,\mathbb{H_C}) :=  \{ X \in \mathbb{H_C}(n) \mid {X}^{\ddagger} = X, \; \text{Tr} \hspace{0.1cm} X = 0 \}. 
    \label{21}
\end{equation}

\subsection{SU$(2,\mathbb{H_C})$  and SU$(3,\mathbb{H_C})$}\label{B}

We developed explicitly two mandatory examples for showing how the formalism at hand works. Then, for $n=2$ we have the $\mathfrak{u}(2,\mathbb{H_C})$; starting from a matrix $2\times2$ on $\mathbb{H_C}$, with arbitrary entries of the form (\ref{01}), the condition of hermiticity (\ref{09}) leads to 
\begin{equation}
    H=\begin{pmatrix}
    \theta_{3}+ij\chi_{3} & \theta_{1}-i\theta_{2}+j\chi_{1}+ij\chi_{2} \\ 
    \theta_{1}+i\theta_{2}-j\chi_{1}+ij\chi_{2} & \theta_{4}+ij\chi_{4}
\end{pmatrix},
\label{h hermitian}
\end{equation}
of the 16 initial real parameters, the imposed restriction leads us to 8 real parameters. The zero-trace condition implicates, $\theta_{3}+ij\chi_{3}=-(\theta_{4}+ij\chi_{4})$, obtaining thus an element of $\mathfrak{su}(2,\mathbb{H_C})$,
\begin{align}
   H=\begin{pmatrix}
    \theta_{3} & \theta_{1}-i\theta_{2} \\
    \theta_{1}+i\theta_{2} & -\theta_{3}
\end{pmatrix}+\begin{pmatrix}
    ij\chi_{3} & j\chi_{1}+ij\chi_{2} \\
   -j\chi_{1}+ij\chi_{2} & -ij\chi_{3}
\end{pmatrix}; 
\label{explicita}
\end{align}
which has been separated into its two sectors. The first part of the structure (\ref{explicita}) is identified with the standar fundamental representation of $\mathfrak{su}(2)$, given by the Pauli matrices $\sigma_{a}$ ($a=1,2,3$). The second sector term can be descomposed in terms of  Pauli-type matrices,  
\begin{equation}
    \tau_1 = \begin{pmatrix} 0 & ij  \\ ij & 0 \end{pmatrix}, \quad 
    \tau_2 = \begin{pmatrix} 0 & j  \\ -j & 0 \end{pmatrix}, \quad 
    \tau_3 = \begin{pmatrix} ij & 0  \\ 0 & -ij \end{pmatrix}. 
    \label{17}
\end{equation}
 Therefore, having explicitly obtained the generators of the group, we proceed to obtain the algebra, which leads to the relations
 \begin{equation}
\begin{gathered}
    [\sigma_{a},\sigma_{b}]=2i\epsilon_{abc}\sigma_{c}, \quad [\tau_{a},\tau_{b}]=-2i {\epsilon}_{abc}\sigma_{c}, \\[2pt]
    [\sigma_{a},\tau_{b}]=2i\epsilon_{abc}\tau_{c}.
\end{gathered}
\label{18}
\end{equation}
with Levi-Civita symbol $\epsilon _{abc}$ as structure constant. We note that the algebra (\ref{18}) has the same structure as $\mathfrak{so}(1,3)$, therefore it can be interpreted that there will be a Lorentz-type symmetry as an internal symmetry.
 Thus, for a four-dimensional background our SU$(2,\mathbb{H_C})$ Yang-Mills theory, will have the same internal and background symmetry.    

For the algebra $\mathfrak{su}(3,\mathbb{H_C})$, we follow the same construction as in the previous case; thereby obtaining a $3 \times 3$ matrix with 16 real parameters, and separating the corresponding Hermitian matrix into its compact and hyperbolic sectors, where the former is identified with the standard fundamental representation of the $\mathfrak{su}(3)$ algebra, spanned by the Gell-Mann matrices $\lambda_a$ ($a=1,2,\ldots,8$)~\cite{Grei,Geor}. As shown in (\ref{explicita}), the second part is the set of  matrices with hyperbolic contributions, which in this case take the following form
\begin{small}
\begin{equation}
\begin{aligned}
\kappa_1 &= \begin{pmatrix} 0 & ij & 0 \\ ij & 0 & 0 \\ 0 & 0 & 0 \end{pmatrix}, \quad
\kappa_2 = \begin{pmatrix} 0 & j & 0 \\ -j & 0 & 0 \\ 0 & 0 & 0 \end{pmatrix}, \quad
\kappa_3 = \begin{pmatrix} ij & 0 & 0 \\ 0 & -ij & 0 \\ 0 & 0 & 0 \end{pmatrix}, \\[6pt]
\kappa_4 &= \begin{pmatrix} 0 & 0 & ij \\ 0 & 0 & 0 \\ ij & 0 & 0 \end{pmatrix}, \quad
\kappa_5 = \begin{pmatrix} 0 & 0 & j \\ 0 & 0 & 0 \\ -j & 0 & 0 \end{pmatrix}, \quad
\kappa_6 = \begin{pmatrix} 0 & 0 & 0 \\ 0 & 0 & ij \\ 0 & ij & 0 \end{pmatrix}, \\[6pt]
\kappa_7 &= \begin{pmatrix} 0 & 0 & 0 \\ 0 & 0 & j \\ 0 & -j & 0 \end{pmatrix}, \quad
\kappa_8 = \frac{1}{\sqrt{3}} \begin{pmatrix} ij & 0 & 0 \\ 0 & ij & 0 \\ 0 & 0 & -2ij \end{pmatrix}.
\end{aligned}
\label{14}
\end{equation}
\end{small}
noting that they exhibit a Gell-Mann-like matrix structure. Computing the commutators, we have
\begin{equation}
\begin{gathered}
    [\lambda_{a},\lambda_{b}]=2i {C}_{abc} \lambda_{c}, \quad [\kappa_{a},\kappa_{b}]=-2i {C}_{abc} \lambda_{c}, \\[2pt]
    [\lambda_{a},\kappa_{b}]=2i {C}_{abc} \kappa_{c}.
\end{gathered}
\label{SU3}
\end{equation}
where $C_{abc}$ turn out to be the structure constants of $\mathfrak{su}(3)$. 

\subsection{The general case SU$(n,\mathbb{H_C})$}
\label{sec:3.3}

We observe that the hyperbolic extensions of the generators in Eqs. (\ref{17}) and (\ref{14}) retain the same structure as the standar complex part; therefore, we can infer the following relationship for the general case between both sectors as reported in \cite{Url05},
\begin{equation}
    \pi_{a}=ij t_{a},
    \label{22}
\end{equation}
where $\pi_a$ correspond to the $\tau_a$-generators in (\ref{17}) and the $\kappa_a$-generators in (\ref{14}); similarly $t_a$ represent the $\sigma_{a}$-generators in the $\mathfrak{su}(2,\mathbb{H_C})$, and $\lambda_a$-generators in the $\mathfrak{su}(3,\mathbb{H_C})$ case. This relationship allows us to show the general rules of algebra, and if the generators are redefined as $T_{a}=\frac{1}{2}t_a$ and $\Pi_{a}=\frac{1}{2}\pi_a$, then we have
\begin{equation}
\begin{gathered}
    [T_{a}, T_b]= i f_{abc} T_c, \quad [\Pi_a, \Pi_b] = -i f_{abc} T_c, \\[2pt]
    [T_a, \Pi_b] = i f_{abc} \Pi_c.
\end{gathered}
\label{12}
\end{equation}
These are the commutation relations of the Lie algebra $\mathfrak{su}(n,\mathbb{H_C})$,where the generators form the base of an R-module\footnote{R denotes a commutative ring with unit element.} \cite{Popov02,Bour}. In the above algebraic relations, it is possible to see that the first relation represents the standard compact $\mathfrak{su}(n)$ algebra; the second relation shows that the $\Pi_a$-generators do not close upon themselves. Note that the last of these relations might imply that $\Pi_a$-generators correspond to an invariant ideal of $\mathfrak{su}(n,\mathbb{H_C})$ \cite{Grei}, but since it is not closed (according to the second relation), the aforementioned possibility cannot be affirmed; in fact, as opposed to $T_a$-generators, the $\Pi_a$-generators it does not form a subalgebra.

The relations in Eq. (\ref{12}) exhibit a structure similar to the doubling of Lie algebras in the TFD formalism \cite{Khan}, where the commutation relations of the generators describing the thermal bath close upon themselves, enabling the description of the thermal symmetries that generate the dissipative evolution \cite{Santa}. 

Once again, Eq. (\ref{22}) enables the calculation of the Killing bilinear form for this group, thereby revealing the metric structure of the space,
\begin{equation}
\begin{gathered}
    \text{Tr} \hspace{0.1cm} (T_{a}T_{b})=\dfrac{1}{2}\delta_{ab}, \quad \text{Tr} \hspace{0.1cm} (\Pi_{a}\Pi_{b})=-\dfrac{1}{2}\delta_{ab}, \\[2pt]
    \text{Tr} \hspace{0.1cm} (T_{a}\Pi_{b})=\dfrac{1}{2}ij\delta_{ab},
\end{gathered}
\label{13}
\end{equation}
noting that, as expected, the metric for the compact sector is positive definite, while the set of $\Pi_a$-generators induces a ne-gative metric confirming the non-compact character of the hyperbolic copy \cite{Geor}. It is noted that the mixed sector contains the Hermitian unit $(ij)$, and thus the metric has not a definite sign. Thus the relations (\ref{13}) imply that we are working in a space with an indefinite metric.

As we have seen, this hyperbolic extension creates a structure similar to SU$(n)$ group, and at level of the Lie algebras can be described through a Cartan decomposition as $\mathfrak{su}(n,\mathbb{H_C})= \mathfrak{su}(n)\oplus ij\mathfrak{su}(n) $. Note that $\mathfrak{so}(1,3)\cong  \mathfrak{sl}(2,\mathbb{C})_{\mathbb{R}}$, where the dimensionality of $\mathfrak{sl}(n, \mathbb{C})_{\mathbb{R}}$ is $2(n^2 - 1)$, which is consistent since a complexification $\mathfrak{su}(n) \otimes_{\mathbb{R}} Z_0$ has been carried out, as can be seen in Eq. (\ref{20}), using the relation (\ref{22}). Consequently, the groups are related as SL$(n, \mathbb{C}) \cong \text{SU}(n, \mathbb{H_C})$, as discussed in \cite{Url05}. 

The idempotent base $(J^{+},J^{-}) $ allows the decomposition of the fundamental algebraic structure of $\mathfrak{su}(n,\mathbb{H_C})$. From (\ref{05}), we have $T_{a}=J^{+}T_{a}^{+}+J^{-}T_{b}^{+}$, and $\Pi_{a}=J^{+}\Pi_{a}^{+}+J^{-}\Pi_{b}^{+}$; although for the $T_{a}$-generators this descomposition is trivial (due to the absence of the $j$-unit), it is not for the hyperbolic sector; in fact $T_{a}^{+}=T_{a}^{-}=T_a$. However, the $\tau_a$-generators, will have the decomposition
\begin{align}
    \tau_1 &= J^+ \begin{pmatrix} 0 & i \\ i & 0 \end{pmatrix} + J^- \begin{pmatrix} 0 & -i \\ -i & 0 \end{pmatrix}, \quad 
    \tau_2 = J^+ \begin{pmatrix} 0 & 1 \\ -1 & 0 \end{pmatrix} + J^- \begin{pmatrix} 0 & -1 \\ 1 & 0 \end{pmatrix}, \nonumber \\[6pt]
    \tau_3 &= J^+ \begin{pmatrix} i & 0 \\ 0 & -i \end{pmatrix} + J^- \begin{pmatrix} -i & 0 \\ 0 & i \end{pmatrix}.
\label{proyecsu2}
\end{align}
and similarly for the $\kappa_a$-generators in (\ref{14}). 
From the properties (\ref{proJ}) and the Eq.(\ref{22}), the decomposition of the entire Lie algebra is derived
\begin{align}
    [T_{a}^{+},T_{b}^{+}]&=if_{abc}T_{c}^{+}, \quad [T_{a}^{-},T_{b}^{-}]=if_{abc}T_{c}^{-}, \nonumber \\ 
    [\Pi_{a}^{+},\Pi_{b}^{+}]&=-if_{abc}T_{c}^{+}, \quad [\Pi_{a}^{-},\Pi_{b}^{-}]=-if_{abc}T_{c}^{-}, \\ 
    [T_{a}^{+},\Pi_{b}^{+}]&=if_{abc}\Pi_{c}^{+}, \nonumber \quad [T_{a}^{-},\Pi_{b}^{-}]=-if_{abc}\Pi_{c}^{-}.
    \label{projection}
\end{align}
The first column of these relations corresponds to the algebras projected onto $J^{+}$, while the second represents the projection of the algebras onto $J^{-}$. Notably, the latter exhibits a sign change, demonstrating that the mixed relations do not decouple into two identical copies. This decomposition will later allow us to identify the $(+, -)$ superscripts with the subsystem and the environment, respectively.

\section{Yang-Mills hypercomplex dynamics}
\label{sec:4}

Given the Lie group defined over $\mathbb{H_C}$, we proceed by acting with this set of linear transformations on a multiplet of hypercomplex fields $\Psi$ as
\begin{equation}
    \Psi \rightarrow U(x)\Psi, \quad \text{where} \quad U(x)=e^{i(\theta^{a}(x)T_{a}+\chi^{a}(x)\Pi_{a})}, 
    \label{23}
 \end{equation}
since our objective is to construct a gauge formalism, the parameters $\{ \theta_{a}, \chi_{a}\}$ are dependent on the spacetime coordinates; this generates local gauge transformations. The hypercomplex field $\Psi$ consists of $n$ components with the following structure
\begin{equation}
    \Psi=\begin{pmatrix}
        \Psi_{1}(x) \\
        \Psi_{2}(x) \\
        \vdots  \\
        \Psi_{n}(x)
    \end{pmatrix}=\begin{pmatrix}
        \psi_{1}(x)+j\phi_{1}(x) \\
        \psi_{2}(x)+j\phi_{2}(x) \\
        \vdots  \\
        \psi_{n}(x)+j\phi_{n}(x)
    \end{pmatrix}; \quad \psi_{i},\phi_{i} \in \mathbb{C}. 
    \label{24}
\end{equation}
Hence, the each component has the form (\ref{01}). Considering the transformation (\ref{23}), the covariant derivative is proposed as follows
\begin{equation}
    D_{\mu}\Psi=\partial_{\mu} \Psi-igC_{\mu}\Psi=\partial_{\mu} \Psi-ig (A^{a}_{\mu}T_{a}+B_{\mu}^{a}\Pi_{a})\Psi, 
    \label{25}
\end{equation}
where $C_{\mu}=A^{a}_{\mu}T_{a}+B_{\mu}^{a}\Pi_{a}$, is the total gauge connection and $g$ is the coupling constant. In (\ref{25}) the $i$ factor is for convenience; in order to obtain the usual scheme in absence of hyperbolic variables. Note that $C_{\mu}\in \mathfrak{su}(n,\mathbb{H_C})$, and it decomposes into a compact part associated with $T_a$, and a hyperbolic (non-compact) part associated with $\Pi_{a}$. It is at this point that a physical identification is given to these gauge fields. In previous works \cite{Car01,Car03}, a scalar field is associated with a subsystem (system of interest) and another with the environment (thermal bath); from this idea, we can propose the notion of assigning the gauge field $A_{\mu}^{a}$ to the subsystem and $B_{\mu}^{a}$ to the environment. The possible application on the gluon-gluon dynamics is discussed in the conclusions. From the relation of generators (\ref{22}), the covariant derivative can be rewritten in terms of a single generating basis
\begin{equation}
    D_{\mu}\Psi=\partial_{\mu}\Psi-ig(A_{\mu}^{a}+ijB_{\mu}^{a})T_{a} \Psi,
    \label{26}
\end{equation}
this allows us to explicitly observe that the gauge field $C_{\mu}$ is a Hermitian quantity and the fields $A_{\mu}^{a},B_{\mu}^{a}\in \mathbb{R}$.  

Considering that the parameters are now dependent on the coordinates and considering the form of the covariant derivative, the total gauge field must have the following transformation law
\begin{equation}
     C^{\prime}_\mu=U(C_{\mu}+\dfrac{i}{g}\partial_{\mu})U^{\ddagger},
     \label{27}
 \end{equation}
 which possesses the same structure as in the $SU(n)$ case, noting that the proposed matrix involution criterion $U^{\ddagger}=U^{-1}$ has now been used. Recall that the structures of interest are the fields $A_{\mu}$ and $B_{\mu}$, hence from (\ref{27}) the decomposition is performed and from infinitesimal transformations, the transformation rules for the fields are obtained; we consider then the infinitesimal transformation  $U\approx1+i\alpha(x)$, where $\alpha(x)=\theta^{a}(x)T_{a}+\chi^{a}\Pi_{a}$, which leads us to the relation $C^{\prime}_\mu=C_{\mu}+\frac{1}{g}\alpha-i[C_{\mu},\alpha]$; where the commutation relations (\ref{12}) have been used, and thus leading to
 \begin{align}
     {A_{\mu}^{a}}^{{\prime}}&=A_{\mu}^{a}+\dfrac{1}{g}\partial_{\mu}\theta^{a}+f_{abc}(A_{\mu}^{b}\theta^{c}-B_{\mu}^{b}\chi^{c}), \label{28} \\
    {B_{\mu}^{a}}^{{\prime}}&=B_{\mu}^{a}+\dfrac{1}{g}\partial_{\mu}\chi^{a}+f_{abc}(A_{\mu}^{b}\chi^{c}+B_{\mu}^{b}\theta^{c}). \label{29}
 \end{align}
 Analyzing the transformation (\ref{28}), it is noted that the first three terms correspond to the standard case; the fourth term $B_{\mu}^{b}\chi^{c}$ of interest as a new contribution, since they are variables used to describe the environment. Therefore, it can be interpreted that a transformation on the subsystem will be perceived by and will transform the environment. The transformation rule for the reservoir(\ref{29}) presents two mixing terms, containing the product between the subsystem-gauge field and the non-compact parameter $A_{\mu}^{b}\chi^{c}$, and the opposite case occurs with the last term $B_{\mu}^{b}\theta^{c}$. This demonstrates the interactive character of the theory, in which the dynamics of the gauge fields will be linked.  
 
 The construction of the curvature tensor will allow us to obtain the dynamics of the total system, through the standard method \cite{Peskin}, by means of the commutator of the covariant derivative,
 \begin{equation}
     [D_{\mu},D_{\nu}]=-igG_{\mu \nu}; \quad G_{\mu \nu}=\partial_{\mu}C_{\nu}-\partial_{\nu}C_{\mu}-ig[C_{\mu},C_{\nu}].
     \label{30}
 \end{equation}
Such as the total gauge field, the curvature tensor decomposes into a compact and a non-compact part, $G_{\mu \nu}=F_{\mu \nu}^{a}T_{a}+W_{\mu \nu}^{a}\Pi_{a}$, where $F_{\mu \nu}^{a}$ corresponds to the field tensor associated with the subsystem or system of interest, and $W_{\mu \nu}^{a}$ is the field tensor of the environment, where the explicit form of these quantities is obtained by decomposing $C_{\mu}$:
 \begin{align}
    F_{\mu \nu}^{a}&=\partial_{\mu}A_{\nu}^{a}-\partial_{\nu}A_{\mu}^{a}+gf_{abc}(A_{\mu}^{b}A_{\nu}^{c}-B_{\mu}^{b}B_{\nu}^{c}), \label{31}\\
    W_{\mu \nu}^{a}&=\partial_{\mu}B_{\nu}^{a}-\partial_{\nu}B_{\mu}^{a}+gf_{abc}(A_{\mu}^{b}B_{\nu}^{c}-A_{\mu}^{b}B_{\nu}^{c}), \label{32}
\end{align}
once again, coupled contributions from the environment on the subsystem are observed, namely, as second-order terms of $B_{\mu}$ in the tensor $F_{\mu \nu}^{a}$, and the mixing of both gauge fields in $W_{\mu \nu}^{a}$.

\subsection{A SU$(n,\mathbb{H_C})$-invariant and equations of motion}
\label{sec:4.1}

We start from the following action, which will allow us to determine the dynamics of the system,
\begin{equation}
    S=-\dfrac{1}{2}\int dx^{n} \: \text{Tr}(G_{\mu \nu}G^{\mu \nu}).
    \label{33}
\end{equation}
as a function of the total dynamics gauge field $S[C_{\mu}]$ will co-rrespond to a SU$(n,\mathbb{H_C})$-invariant and $n$ is the dimension of an arbitrary background. At first glance, the Lagrangian has the same structure as in the classical Yang-Mills case, but the decomposition of the curvature tensor $G_{\mu \nu}$ will allow us to know the true structure of this invariant:  
\begin{align}
    \mathcal{L}&=-\dfrac{1}{2}\text{Tr}((F_{\mu \nu}^{a}T_{a}+W_{\mu \nu}^{a}\Pi_{a})(F^{\mu \nu}_{a}T^{a}+W^{\mu \nu}_{a}\Pi^{a}))\nonumber \\
    &=-\dfrac{1}{2}\text{Tr}(F_{\mu \nu}^{a}F^{\mu \nu}_{a}T_{a}T^{a})-\dfrac{1}{2}\text{Tr}(F_{\mu \nu}^{a}W^{\mu \nu}_{a}T_{a}\Pi^{a}) \nonumber \\
    &-\dfrac{1}{2}\text{Tr}(W_{\mu \nu}^{a}F^{\mu \nu}_{a}\Pi_{a}T^{a})-\dfrac{1}{2}\text{Tr}(W_{\mu \nu}^{a}W^{\mu \nu}_{a}\Pi_{a}\Pi^{a})\nonumber \\ 
    &=-\dfrac{1}{4}F_{\mu \nu}^{a}F^{\mu \nu}_{a}+\dfrac{1}{4}W_{\mu \nu}^{a}W^{\mu \nu}_{a}-\dfrac{1}{2}ijF_{\mu \nu}^{a}W^{\mu \nu}_{a}. \label{34}
\end{align}
Therefore, defining a Lagrangian over the commutative ring $\mathbb{H_C}$ leads to a Hermitian form, which agrees with the algebraic structure that extends the notion of scalar quantities. The equations of motion can be found by performing the variation of this Lagrangian (\ref{34}), which would lead to tedious calculations, but taking advantage of the action (\ref{33}), the general equation of motion can be obtained \cite{Tong}, $D^{\mu}G_{\mu \nu}=\partial^{\mu}G_{\mu \nu}-ig[C^{\mu},G_{\mu \nu}]=0$, basically by considering the full gauge field $C_{\mu}$ as the dynamical variable, whereupon we proceed with the decomposition into the basis of generators, obtaining the following equations of motion:
\begin{align}
    \partial^{\mu} F_{\mu \nu}^{a} & +gf_{abc}(A_{b}^{\mu}F_{\mu \nu}^{c}-B^{\mu}_{b}W_{\mu \nu}^{c})=0, \label{35} \\
    \partial^{\mu} W_{\mu \nu}^{a} & +gf_{abc}(A_{b}^{\mu}W_{\mu \nu}^{c}+B^{\mu}_{b}F_{\mu \nu}^{c})=0, \label{36}
\end{align}
where the first of these (\ref{35}) belongs to the equation of motion for the subsystem, and (\ref{36}) describes the dynamics of the environment. Note that the full system represents a system of highly non-linear coupled differential equations, which is not solvable in a general way. By virtue of the elements provided by the commutative ring, through the idempotent basis $(J^{+},J^{-})$, another representation of the equations of motion can be reached starting from the algebraic structure proposed in \cite{Car01}; therefore we decompose the total gauge field $C_{\mu}$ in the following manner, see Eq. (\ref{05}),
\begin{equation}     C_{\mu}=J^{+}C_{\mu}^{+}+J^{-}C_{\mu}^{-}; \quad C_{\mu}^{\pm}=(A_{\mu}^{a}\pm iB_{\mu}^{a})T_{a}, 
\label{37}
\end{equation}
Noting that $C_{\mu}^{\pm}$ is been projected onto the compact basis of generators (one is free to choose which basis to project onto). This allows us to rewrite the curvature tensor in terms of two fields %($C_{\mu}^{\pm}$):
\begin{align}
    G_{\mu \nu}=J^{+}(\partial_{\mu}C_{\nu}^{+}&-\partial_{\nu}C_{\mu}^{+}-ig[C_{\mu}^{+},C_{\nu}^{+}])\nonumber \\
    &+J^{-}(\partial_{\mu}C_{\nu}^{-}-\partial_{\nu}C_{\mu}^{-}-ig[C_{\mu}^{-},C_{\nu}^{-}]), \label{38}
\end{align}
where the orthogonality property of the idempotents $J^{+}J^{-}=0$ has been used. Summarizing, the curvature field tensor of the system is now decomposed in the new basis as
\begin{equation}
    G_{\mu \nu}=J^{+}G_{\mu \nu}^{+}+J^{-}G_{\mu \nu}^{-}; \quad G_{\mu \nu}^{\pm}=\partial_{\mu}C_{\nu}^{\pm}-\partial_{\nu}C_{\mu}^{\pm}-ig[C_{\mu}^{\pm},C_{\nu}^{\pm}].
    \label{39}
\end{equation}
Allowing the decoupling of the equations of motion as follows
\begin{align}
    D^{\mu}G_{\mu \nu}=J^{+}(\partial_{\mu}G_{\mu \nu}^{+}&-ig[C^{\mu +},G_{\mu \nu}^{+}]) \nonumber \\
    +&J^{-}(\partial_{\mu}G_{\mu \nu}^{-}-ig[C^{\mu -},G_{\mu \nu}^{-}])=0.
    \label{40}
\end{align}
Then, projecting into the idempotent basis, 
\begin{equation}
    \partial_{\mu}G_{\mu \nu}^{+}-ig[C^{\mu +},G_{\mu \nu}^{+} ]=0, \quad \partial_{\mu}G_{\mu \nu}^{-}-ig[C^{\mu -},G_{\mu \nu}^{-}]=0.
\end{equation}

Note that, there is no coupling between and $C_{\mu}^{+}$ and $C_{\mu}^{-}$. We having an equation of motion associated with each basis, interpreting that the solution for $C_{\mu}^{+}$ corresponds to the system of interest, and $C_{\mu}^{-}$ associated with the environment. Therefore, the equations of motion in the idempotent basis decouple the fields, yielding a simpler and more tractable structure than Eqs. (\ref{35}) and (\ref{36}), although the concept of field mixing remains.

\section{Concluding remarks}
\label{sec:con}
A non-Abelian field theory has been formulated over a commutative ring, where local gauge transformations are are composed of elements of the form (\ref{01}).By itself, this algebraic structure enables the expansion of the internal degrees of freedom, consequently doubling the dimensionality of the Lie algebra. This yields two distinct sectors: one composed of compact generators and another of non-compact generators derived from the hyperbolic extension, as shown in Eq. (\ref{20}). Now, quantities such as the gauge fields and the Lagrangian of the theory are elements of the Hermitian subset (\ref{07}). This formalism also enables the coupling constant $g$ to assume values in the subset $Z_0$, yielding the form $g = g_1 + ij g_2$. Here, $g_1$ is interpreted as the coupling of the subsystem, while $g_2$ represents the coupling of the environment. This generalization will expand the dynamical expressions derived in this manuscript, a feature that will be addressed in future investigations. Based on the developments in \ref{B}, it will be possible to study bipartite systems through the SU$(2, \mathbb{H_C})$ group. Likewise, it enables the exploration of gluon-gluon dynamics, which, upon coupling to matter, will lead to the study of the quark-gluon plasma via SU$(3, \mathbb{H_C})$. The coupling between both fields occurs at the dynamical level, as observed in Eqs. (\ref{35}) and (\ref{36}). However, employing the $(J^{+}, J^-)$ basis is an advantage of the ring structure, as it allows for the decoupling of the equations of motion, yielding two copies of the Yang-Mills equations but for complex gauge fields. This enables the derivation of analytical solutions using the Wu-Yang ansatz to obtain 't Hooft-Polyakov monopoles \cite{Actor}. Within these solutions for $C^{\pm}$, the dissipative behaviors will manifest by the presence of decaying/increasing modes, such as  the scalar field frameworks \cite{Car03,Car01}. Consequently, the dynamics within the idempotent framework yields exact solutions for dissipative fields interacting with thermal baths.

A Hamiltonian analysis of the action in Eq. (\ref{33}) is left for future work. This will lead us to explore the constraints of the theory in order to derive the dynamical algebra. Comparing these results with the analysis of the standard theory \cite{Hanson}, will allow us to observe how the structures are deformed due to the dissipative dynamics. For a subsequent quantization of the theory, the Faddeev-Jackiw formalism is more convenient due to its Lagrangian approach \cite{Bar} and its straightforward implementation into the path integral, thereby yielding the partition function \cite{Toms}.
 Analyzing its topological framework will enable us to probe the non-perturbative regime of the theory; determining the $\theta$-term will elucidate the vacuum structure of both the subsystem and the environment, where instanton solutions would describe the relationship between these two vacua \cite{Tong}. As demonstrated, the hypercomplex extension of Yang-Mills theory provides an avenue to explore a broad spectrum of problems within the context of TFD.
 
%\section*{Data Availability} 
%No data associated with the manuscript.

%\section*{Declaration of competing interest} 
%The authors declare that they have no known competing financial interests or personal relationships that could have appeared to influence the work reported in this paper.

\section*{Acknowledgements}
We thank B. R. López Raymundo for his discussions on this work. This work was supported by the Sistema Nacional de Investigadores (M\'exico). C. M. Lopez Arellano would like to thank Secretariat of Science, Humanities, Technology and Innovation (Secihti, M\'exico) for financial support (CVU: 2043680, 2024-2026).

\bibliographystyle{unsrtnat}
\bibliography{bib}

\end{document}